\documentclass[preprint,
showpacs,preprintnumbers,amsmath,amssymb]{revtex4}

\usepackage{graphicx,subfigure}
\usepackage{lscape,graphicx}
\usepackage{dcolumn}
\usepackage{bm}
\usepackage{amsmath}

\begin{document}
\preprint{UAB-FT-609}
\title{Imprint of spatial curvature on inflation power spectrum }
\author{ Eduard  Mass{\'o}$^a$,  Subhendra  Mohanty$^{a,b}$, Akhilesh Nautiyal$^b$ \& Gabriel Zsembinszki$^a$}%
\affiliation{$^a$ Department of Physics \& IFAE, Universitat
Aut{\`o}noma de Barcelona, 08193 Bellaterra, Spain\\ $^b$  Physical
Research Laboratory, Ahmedabad 380009, India}
\date{\today}%

\begin{abstract}
If the universe had a large curvature before inflation there is a
deviation from the scale invariant perturbations of the inflaton
at the beginning of inflation. This may have some effect on the
CMB anisotropy at large angular scales.
  We calculate the density perturbations for both open and
closed universe cases using the Bunch-Davies vacuum condition on
the initial state.
 We use our power spectrum to
calculate the temperature anisotropy spectrum and compare the results with the WMAP three
year data. We find that our power spectrum gives a lower quadrupole anisotropy
when $\Omega-1 >0$, but matches the temperature anisotropy
calculated from the standard Ratra-Peebles power spectrum at
large $l$. The determination of spatial curvature from
temperature anisotropy data is not much affected by the
different power spectra  which arise from the choice of different boundary conditions for the inflaton perturbation.
\end{abstract}


\maketitle
\section{Introduction}

The curvature of the universe goes down exponentially after
the start of inflation \cite{guth}. If there is a residual
curvature
still present by the time the scales which are entering our
horizon at present were leaving the inflationary horizon there
will be deviation from the scale invariant perturbations  due to
non-zero curvature. The corrections to the power spectrum at
horizon scales are multiplicative powers of
$(1\pm K/\beta^2)$,
 where the curvature $K=(\Omega_0-1)(a_0 H_0)^2$ and
$\beta$ is the comoving canonical wavenumber. We calculate the
primordial power spectrum for the case closed and open universe at
the time of inflation. We choose the Bunch-Davies boundary
condition to normalize the wave-functions. For the case of closed universe we obtain the following expression for the power spectrum
\begin{equation}
P_{\cal R}(\beta)=\frac{H_\lambda^4}{2 \pi^2 \dot
\phi^2}\frac{1}{\left(1+\frac{K}{\beta^2}\right)^2},\quad \quad
\frac{\beta}{\sqrt{K}}=3,4,5 \cdots \,\,( {\rm for}\ K>0)
 \label{closed1}
\end{equation}
and for the case of inflation in an open universe
\begin{equation}
P_{\cal R}(\beta)=\frac{H_\lambda^4}{2 \pi^2 \dot
\phi^2}\frac{1}{\left(1-\frac{|K|}{\beta^2}\right)^2
\left(1+\frac{|K|}{\beta^2}\right)},\quad \quad
\frac{\beta}{\sqrt|K|}>1, \,\,( {\rm for} \ K<0) \label{open1}
\end{equation}
where $\phi$ is the inflaton field. In the case of closed universe $\beta$ takes discrete
values in units of $\sqrt{K}=R_c^{-1}$ ($R_c$ being the curvature
radius), the modes corresponding to $\beta/\sqrt{K}=1,2$ can be
eliminated by gauge transformations \cite{abbott} so there is a
large-wavelength cutoff at $\beta_c^{-1}= R_c/3$. This large
wavelength cut-off in a closed universe has been used to explain
the observed low CMB anisotropy at low multi-poles \cite{efsth}
and \cite{lasenby}. In the case of open universe only modes with
$\beta > \sqrt{|K|}$ cross the inflationary horizon.

Our result for the power spectrum in the closed and open universe
cases differs from the phenomenological power spectrum
\cite{Hu1997},
\begin{equation}
P_{\cal R}(\beta) = \frac{H_\lambda^4}{2 \pi^2 \dot \phi^2}
\,\,\,\frac{1} {1+\frac{K}{\beta^2}} \label{HSWZ}
\end{equation}
used in the calculation of CMB anisotropies in both the closed and
open cases. Our results agree qualitatively with (\ref{HSWZ}) in
that the power at small $\beta$ is suppressed in the closed
universe inflation (\ref{closed1}) and enhanced in the open
universe (\ref{open1}).

According to inflation \cite{guth}, the curvature of the present
universe $\Omega_0-1$ is related to the curvature at any time
during  inflation $\Omega_{i}-1$ as
 \begin{equation}
  \frac{\Omega_0-1}{\Omega_i-1}=\left(\frac{a_i H_i}{a_0
  H_0}\right)^2.
 \end{equation}
If $a_i$ is the scale factor at the time during inflation when
scales of the size of our present horizon were exiting the
inflationary horizon then $a_0 H_0= a_i H_i$ and
$\Omega_0=\Omega_i$. If at the beginning of
inflation $(\Omega_{start}-1)=O(1)$ then in order to have a
deviation of say one-percent from unity in the present curvature,
the number of e-foldings prior to the $a_i$ must be small. Putting
an upper bound on the present curvature $(\Omega_0-1)$ from
observations also puts a lower bound on the number of extra
e-foldings necessary in inflation in addition to the minimum
number needed to solve the horizon problem \cite{ellis}.

The geometry of the universe can be determined from the CMB
anisotropy from the angular size of the acoustic peak. However the
constraints on the density of the universe $\Omega$ depend upon
priors like the value $H_0$ and $\Omega_\lambda$. For example the
combination of WMAP \cite{wmap3} , LSS \cite{Tonry:2003zg} and HST \cite{Freedman:2000cf} supernovae observations  gives a
constraint on the density of the universe as
$(\Omega-1)=0.06^{+0.02}_{-0.02}$ \cite{Dunkley} which means that
the curvature at one-$\sigma$ could be as large as $K/(a_0
H_0)^2=0.08$. In the case of the closed universe the power
spectrum (\ref{closed1}) $P_{\cal R} \propto (1+K/\beta^2)^{-2}$
at the scale of the horizon $\beta=a_0 H_0$ would be suppressed by about
$16\%$  compared to the power for the flat universe.

We use our power spectrum to
calculate the temperature anisotropy spectrum and compare the results with the WMAP three
year data. We find that our power spectrum gives a lower quadrupole anisotropy
when $\Omega-1 >0$, but matches the temperature anisotropy calculated from the standard Ratra-Peebles power spectrum at large $l$. We also find that using the closed universe power spectrum (\ref{closed2})  for larger  values of $ \Omega_0$ the quadrupole anisotropy is suppressed more and fits the WMAP data better .  This supports the idea proposed in \cite{efsth} that a positive spatial curvature should suppress the power at low $l$.
\section{Scalar power spectrum }

We expand the inflaton  field  $\phi({\bf x},t)\equiv \phi(t)+
\delta \phi({\bf x},t)$, where the perturbations $\delta \phi$
around the constant background $\phi(t)$ obey the minimally
coupled KG equation
\begin{equation}
\ddot {\delta \phi} + 3 \frac{\dot a}{a} \dot {\delta\phi}
-\frac{1}{a^2} \nabla^2 \delta\phi=0.
\end{equation}
With the separation of variables
\begin{equation}
\delta \phi({\bf x} ,t)= \sum_k  \delta\phi_k(t)Q({\bf x},k)
\end{equation}
the KG equation can be split as
\begin{eqnarray}
\ddot {\delta \phi_k} + 3 \frac{\dot a}{a} \dot {\delta \phi_k}
+\frac{k^2}{a^2}
\delta \phi_k=0\\[8pt]
 \nabla^2 Q({\bf x},k)=-k^2 Q({\bf x},k) \label{Q}
\end{eqnarray}
where $\nabla^2$ is the Laplacian operator for the spatial part.
Making the  transformation $d \eta = dt/a $ and
$\sigma(\eta,k)=a(\eta)\delta \phi_k(\eta)$ we get the KG equation
for $\sigma(\eta,k)$
\begin{equation}
\sigma^{\prime \prime} + (k^2-\frac{a^{\prime \prime}}{a})\sigma=0
\label{KGa}
\end{equation}
where primes denote derivatives w.r.t conformal time $\eta$.


 The Friedman equations in conformal time are,
\begin{eqnarray}
 \left( \frac{a^{\prime}}{a} \right)^2 &=&
\frac{8 \pi G}{3} \rho\, a^2 -K \label{friedman-1}
\\[8pt]
\left( \frac{a^{\prime}}{a} \right)^\prime &=& -\frac{4 \pi G}{3}
(\rho + 3 p) a^2. \label{friedman-2}
\end{eqnarray}
Consider the universe with cosmological constant and curvature, then
$ \rho=\rho_\lambda$ and $p=-\rho_\lambda$ and we get using the
Friedman equations,
\begin{eqnarray}
\frac{a^{\prime \prime}}{a}=\frac{16 \pi G}{3} \rho_\lambda \,a^2 -K
 \equiv 2 a^2 H_\lambda^2 -K
 \label{a2prime}
\end{eqnarray}
where $H_\lambda=(\frac{8 \pi G}{3} \rho_\lambda)^{1/2}$ is the
Hubble parameter during pure inflation. Substituting (\ref{a2prime})
in the KG equation (\ref{KGa}) we obtain,
\begin{equation}
\sigma^{\prime \prime} + (k^2-2 a^2 H_\lambda^2 +K)\sigma=0.
\label{KG}
\end{equation}
The curvature affects the wave equation of $\sigma(\eta)$ in the
explicit dependence $K$ and also in the changed dynamics of
$\eta-$dependence of the scale factor $a$ which is important in
the early stages of inflation.

The scalar field perturbation can be written as
\begin{equation}
\delta \phi({\bf x},\eta)=\sum_k \,\frac{\sigma(\eta,k)}{a(\eta)}
\,Q({\bf x},k) \label{Qexpand}
\end{equation}
where $\sigma(\eta)$ is the solution of equation (\ref{KG}) and
the spatial harmonics $Q({\bf x},k) $ are solutions of equation
(\ref{Q}) \cite{abbott}. One can separate the radial and angular
modes of $Q_{\beta}^{lm}(r,\theta,\phi)$ as
\begin{equation}
Q_{\beta}^{lm}(r,\theta,\phi)=\Phi_{\beta}^{l }(r)\,Y_l^m(\theta,
\phi)
\end{equation}
where $\beta=(k^2 +K)^{1/2}$ are the eigenvalues of the
radial-part of the Laplacian with eigenfunctions given by the
hyperspherical Bessel functions $\Phi_{\beta}^{l }(r)$ which are
listed in \cite{abbott}. In the limit $K \rightarrow 0$, the
radial eigenfunctions $\Phi_{\beta}^{l } (r) \rightarrow
j_l(k\,r)$. The main properties that are needed for the
calculation of the power spectrum are orthogonality
\begin{equation}
\int \gamma r^2drd\Omega
Q_{\beta}^{lm}(r,\theta,\phi)Q_{\beta'}^{*\,l'm'}(r,\theta,\phi)
=\frac{1}{\beta^2}\delta_{ll'}\delta_{mm'}\delta_{\beta\beta'}
\label{ortho1}
\end{equation}
where $\gamma=(1+\frac{Kr^2}{4})^{-3}$ is the determinant of the
spatial metric,
and completeness
\begin{equation}
\sum_{l,m} \int \beta^2 d\beta Q_{\beta}^{lm}(r,\theta,\phi)
Q_{\beta}^{*\,lm}(r',\theta',\phi')=\gamma^{-1} \frac{1}{r^2}
\delta (r-r')\delta(\theta-\theta') \delta(\phi-\phi').
\end{equation}
In case of closed universe the integral over $\beta$ is replaced
by sum over the integers $\beta/\sqrt{K}=3,4,5...$. For open and
flat universes $\beta$ is a real non-negative variable.

The gauge invariant perturbations are a combination of metric
and inflaton perturbations. The curvature perturbations are gauge
invariant and at super-horizon scales are related to the inflaton
perturbations as
\begin{equation}
{\cal R}({\bf x},\eta) = \frac{H}{\dot \phi} \delta \phi({\bf
x},\eta). \label{R}
\end{equation}
Curvature perturbations generated during inflations are frozen
outside the horizon till they re-enter in the radiation or matter
era. CMB anisotropies at large angles are caused by curvature
perturbations in the surface of last scattering which enter in the
matter era. The Sachs-Wolfe effect at large angles, relates the
temperature perturbation in the direction ${\bf\hat n}$ observed
by the observer located at the point $({\bf x_0},\eta_0)$ to the
curvature perturbation at the point $({\bf x}_{LS},\eta_{LS})$ in
the
 LSS,
 \begin{equation}
 \frac{\delta T({\bf x_0,\hat n},\eta_0)}{T}= \frac{1}{5}
 {\cal R}({\bf x}_{LS},\eta_{LS})
 \label{Tx}
 \end{equation}
where ${\bf x}_{LS}={\bf \hat n} (\eta_{LS}-\eta_0)$. Using the
completeness of $Q_{\beta}^{lm}(r,\theta,\phi)$  we can expand
${\cal R}$ as a sum-over the eigenmodes,
\begin{equation}
 {\cal R}({\bf x}_{LS},\eta_{LS})=\sum_{l m} \int \beta^2
 d\beta \,\left[\frac{H}{\dot \phi}\,
 \delta \phi_\beta(\eta)\right]_{\eta=\eta_*}
\,Q_{\beta}^{lm}({\bf x}_{LS}). \label{RQ}
\end{equation}
Here we have used the fact that ${\cal R}$ does not change after
exiting the horizon during inflation (at a conformal time which we
shall denote by $\eta_*$) till it re-enters the horizon close to
the LS era. Using the Sachs-Wolfe relation (\ref{Tx}) and the mode
expansion of the curvature perturbation (\ref{RQ}) and using the
orthogonality (\ref{ortho1}) of $Q_{\beta}^{lm}$, we obtain
\begin{equation}
\left\langle \frac{\delta T({\bf \hat n_1)}}{T}\frac{\delta T({\bf
\hat n_2)}}{T}\right\rangle = \sum_l  \frac{2l +1}{4
\pi}\,P_l({\bf \hat n_1}\cdot {\bf \hat n_2)} \int \beta^2 d \beta
\frac{1}{25} |{\cal{R}}(\beta, \eta_*)|^2 \,
|\Phi_{\beta}^l(\eta_{0}-\eta_{LS})|^2 \label{Tn}.
\end{equation}
The angular spectrum  $C_l$ of temperature anisotropy defined by
\begin{equation}
\left\langle \frac{\delta T({\bf \hat n_1)}}{T}\frac{\delta T({\bf
\hat n_2)}}{T} \right\rangle=\sum_l  \frac{2l +1}{4 \pi}\,P_l({\bf
\hat n_1}\cdot {\bf \hat n_2)} \, C_l \label{Cl1}
\end{equation}
can be written in terms of the power spectrum of curvature
perturbations by comparing (\ref{Cl1}) with (\ref{Tn}),
\begin{equation}
C_l = 4\pi\int \frac{d \beta}{\beta} \frac{1}{25} |P_{\cal
R}(\beta)|^2 \, |\Phi_\beta^l(\eta_0-\eta_{LS})|^2
\end{equation}
where the power spectrum of curvature perturbations is defined as
\begin{equation}
P_{\cal R}(\beta)=\frac{\beta^3}{2 \pi^2} \left[\left(\frac{H}{\dot
\phi}\right)^2 |\delta\phi_\beta(\eta)|^2\right]_{\eta=\eta_*}.
\end{equation}

We shall now derive the power spectrum for the open and closed
inflation universes.

\section{Closed universe inflation} For a closed
universe, $K > 0$, from the Friedman equation (\ref{friedman-1})
we get
\begin{equation}
\dot{a}= H_\lambda a\sqrt{1-\frac{K} {H_{\lambda}^2a^2}}
\end{equation}
which can be integrated to give
\begin{equation}
a(t)= \frac{\sqrt{K}}{H_\lambda}\cosh{H_\lambda t}.
\end{equation}
We consider as initial conditions the moment when the inflaton
energy density starts to dominate over the curvature, i.e. for
$t=0$ we have $a(0)=\frac{\sqrt{K}}{H_\lambda}$.  The conformal
time is then given by
\begin{eqnarray}
\eta(a)  =
 \frac{-1}{\sqrt{K}}\arcsin\frac{\sqrt{K}}{a H_\lambda}.
\label{etaK2}
\end{eqnarray}
The conformal time spans the interval
$\eta=(-\frac{\pi}{2\sqrt{K}},0)$ as the scale factor $a$ varies
between $(\frac{\sqrt{K}}{H_\lambda},\infty)$, so for $K\neq 0$
our initial conditions are different from the standard inflation
case. The dependence of the scale factor on the conformal time is
obtained from (\ref{etaK2})
\begin{equation}
a(\eta)=-\frac{\sqrt{K}}{H_\lambda} \frac1{\sin\sqrt{K}\eta}.
\end{equation}
The conformal time KG equation (\ref{KG}) for the
closed-inflationary universe is of the form
\begin{equation}
\sigma^{\prime \prime}(\eta) + \left[k^2-K\left(2 {\rm cosec}^2
\sqrt{K}\eta-1\right)\right]\sigma(\eta) =0. \label{KGK1}
\end{equation}
This equation can be solved  exactly and the solutions are
\begin{eqnarray}
\sigma(\eta) & = & c_1 \left(-\sqrt{K}\cot\sqrt{K}\eta
+i\sqrt{k^2+K} \right)
e^{i\sqrt{k^2+K}\eta}\nonumber\\[8pt]
&+&  c_2 \left(-\sqrt{K}\cot\sqrt{K}\eta -i\sqrt{k^2+K} \right)
e^{-i\sqrt{k^2+K}\eta}.\label{sigma1}
\end{eqnarray}
The normalization constants $c_1$ and $c_2$  are determined by
imposing the Bunch-Davies vacuum condition i.e the assumption that
in the infinite past limit $\eta \rightarrow -\pi/(2{\sqrt K})$,
$\sigma$ is a plane wave which obeys the canonical commutation
relation
  \begin{equation}
  \sigma^* \sigma'-{\sigma^*}'\sigma = i
  \label{com1}
 \end{equation}
from which we obtain
\begin{equation}
|c_1|=\frac{1}{\sqrt{2}(k^2+K)^{3/4}},\quad c_2=0.
\end{equation}
Replacing these constants into (\ref{sigma1}) we find that in the
limit $\eta \rightarrow -\pi/(2\sqrt{K})$,
\begin{equation}
\sigma(\eta \rightarrow -\pi/(2\sqrt{K}))\equiv \sigma_{BD}=
\frac{1}{\sqrt{2 \beta}} e^{i \beta \eta} \label{bd}
\end{equation}
where
\begin{equation}
\beta=(k^2+K)^{1/2}.
\end{equation}
We will assume that the vacuum state of the universe $|0\rangle$
is the state in which there are no $\sigma_{BD}$ particles.  The
creation and annihilation operators are for the the BD vacuum can
be written as
\begin{equation}
\sigma_{BD}({\bf x},\eta)= \sum_{l m} \int \beta^2 d\beta
\left(a_{\beta l m}\, Q_\beta^{lm}({\bf x})\,\frac{e^{i\beta
\eta}}{\sqrt{2 \beta}} \,\,+\,\, a_{\beta l m}^\dagger\,
{Q^*}_\beta^{lm}({\bf x})\,\frac{e^{-i\beta \eta}}{ \sqrt{2
\beta}} \right).
\end{equation}
Using the commutation relation (\ref{com1}) of $\sigma_{BD}$ and
the orthogonality of $Q_{\beta}^{lm}$ (\ref{ortho1}) we see that
the creation and annihilation operators obey the canonical
commutation relations
\begin{equation}
[a_{\beta l m} , a^\dagger_{\beta^\prime l^\prime m^\prime}]= \,
\frac{1}{\beta^2} \delta(\beta-\beta^\prime)\delta_{l l^\prime}
\delta_{m m^\prime} .
\end{equation}
>From the foregoing discussion it is clear that $\beta$ is the
radial canonical momentum. The quantum fluctuations become
classical when $\beta=a H$. We shall evaluate the power spectrum
at horizon crossing, as the modes do not change after exiting the
inflation horizon till they re-enter the horizon in the radiation
or matter era.

Substituting the constants $c_1$ and $c_2$ in the general solution
(\ref{sigma1}) and going back to the $\delta \phi$, we find that
\begin{equation}
\langle 0|\delta\phi_\beta(\eta)^2|0\rangle=\frac{1}{a(\eta)^2}
\left[ \frac{\beta^2+K \cot^2\sqrt{K} \eta}{2 \beta^{3}}\right].
\end{equation}
We want to evaluate the  spectrum of perturbations at
horizon crossing. The horizon crossing condition is given by
\begin{equation}
\beta=a_* H(a_*)= a_*
\left(H_\lambda^2-\frac{K}{a_*^2}\right)^{1/2}
\label{horizon_crossing1}
\end{equation}
 from which we obtain the values of the scale factor
\begin{equation}
a_*=\frac{(\beta^2+K)^{1/2}}{H_\lambda}
\end{equation}
and of the conformal time
\begin{equation}
\eta_*=-\frac1{\sqrt{K}}\arctan {\frac{\sqrt{K}}{\beta}}
\end{equation}
at horizon crossing. The corresponding value of the Hubble
parameter is
\begin{equation}
H(a_*)= H_\lambda \frac{\beta}{(\beta^2+K)^{1/2}}.
\end{equation}
The power spectrum ${\cal P}(\beta)$ in this case is given by
\begin{equation}
P_{\cal R}(\beta)=\frac{H_\lambda^4}{2 \pi^2 \dot
\phi^2}\frac{1}{\left(1+\frac{K}{\beta^2}\right)^2}.
 \label{closed2}
\end{equation}

\section{Open universe inflation} Now we consider the
case of an open universe with $K<0$. From the Friedman equation
(\ref{friedman-1})
 we have
\begin{equation}
\dot a=H_\lambda a\sqrt{1+\frac{|K|} {H_{\lambda}^2a^2}}
\end{equation}
where we work with the absolute value of the curvature, taking
into account that $|K|=-K$ in this case. The above expression can
be integrated to give
\begin{equation}
a(t)= \frac{\sqrt{|K|}}{H_\lambda}\sinh{H_\lambda t}
\end{equation}
with initial condition $a(t=0)=0$. The conformal time is
\begin{eqnarray}
\eta(a)
= \frac{-1}{\sqrt{|K|}}{\rm arcsinh}\frac{\sqrt{|K|}}{H_\lambda a}.
\label{etaK1}
\end{eqnarray}
The conformal time spans the interval $\eta=(-\infty, 0)$ as the
scale factor varies in the interval $a=(0, \infty)$. We can solve
for $a(\eta)$ and obtain
\begin{equation}
a(\eta)=-\frac{\sqrt{|K|}}{H_\lambda} \frac1{\sinh\sqrt{|K|}\eta}.
\end{equation}
The conformal time KG equation for the open-inflationary universe is
of the form
\begin{equation}
\sigma^{\prime \prime}(\eta )+ \left[k^2-|K|\left(2\,{\rm
cosech}^2 \sqrt{|K|}\eta+1 \right) \right]\sigma(\eta)=0.
\label{KGK-1}
\end{equation}
This equation has exact solutions
\begin{eqnarray}
\sigma(\eta) & = & c_1\left(-\sqrt{|K|}\coth\sqrt{|K|}\eta+
i\sqrt{k^2-|K|}\right)e^{i\sqrt{k^2-|K|}\,\eta}\nonumber\\
& + & c_2\left(\sqrt{|K|}\coth\sqrt{|K|}\eta+
i\sqrt{k^2-|K|}\right)e^{-i\sqrt{k^2-|K|}\,\eta}. \label{etaK-1}
\end{eqnarray}
The normalization constants $c_1$ and $c_2$ are chosen so that in
the infinite past $\eta \rightarrow - \infty$ limit one gets plane
waves satisfying the following relation
  \begin{equation}
  \sigma^* \sigma'-{\sigma^*}'\sigma = i
 \end{equation}
and obtain,
\begin{equation}
|c_1|=\frac{1}{\sqrt{2}k (k^2-|K|)^{1/4}},\,\quad c_2=0.
\end{equation}
We then obtain for the magnitude of
$\delta\phi_\beta(\eta)=\sigma(\eta)/a(\eta)$ the expression,
\begin{equation}
|\delta\phi_\beta(\eta)|^2=\frac{1}{a(\eta)^2}\,\left[\frac{\beta^2
+ |K|\coth^2\sqrt{|K|}\eta}{2 (\beta^2+|K|)\, \beta }\right].
\end{equation}
where for the open universe,
\begin{equation}
\beta=(k^2-|K|)^{1/2}.
\end{equation}
The horizon crossing condition is given by
\begin{equation}
\beta=a_* H(a_*)= a_*
\left(H_\lambda^2+\frac{|K|}{a_*^2}\right)^{1/2}
\label{horizon_crossing2}
\end{equation}
 and we obtain for the scale factor at Hubble crossing
\begin{equation}
a_*=\frac{(\beta^2-|K|)^{1/2}}{H_\lambda}
\end{equation}
and the corresponding conformal time is given by
\begin{equation}
\eta_*=-\frac1{\sqrt{|K|}}{\rm arctanh}\frac{\sqrt{|K|}}{\beta}.
\end{equation}
The Hubble parameter at horizon crossing is
\begin{equation}
H(a_*)=H_\lambda \frac{\beta}{\sqrt{\beta^2-|K|}}.
\end{equation}
We notice that in an open-universe stage of inflation, only the
modes satisfying the condition $\beta^2> |K|$ will cross the
Hubble radius.

With this, we obtain the following expression for the curvature
power spectrum at Hubble crossing
\begin{equation}
P_{\cal R}(\beta)=\frac{H_\lambda^4}{2 \pi^2 \dot
\phi^2}\frac{1}{\left(1-\frac{|K|}{\beta^2}\right)^2
\left(1+\frac{|K|}{\beta^2}\right)}. \label{open2}
\end{equation}

\section{Effect of curvature on temperature anisotropy spectrum}
\begin{figure}[htb]
\begin{center}
\includegraphics[width=16cm, height=16.5cm,angle=-90]{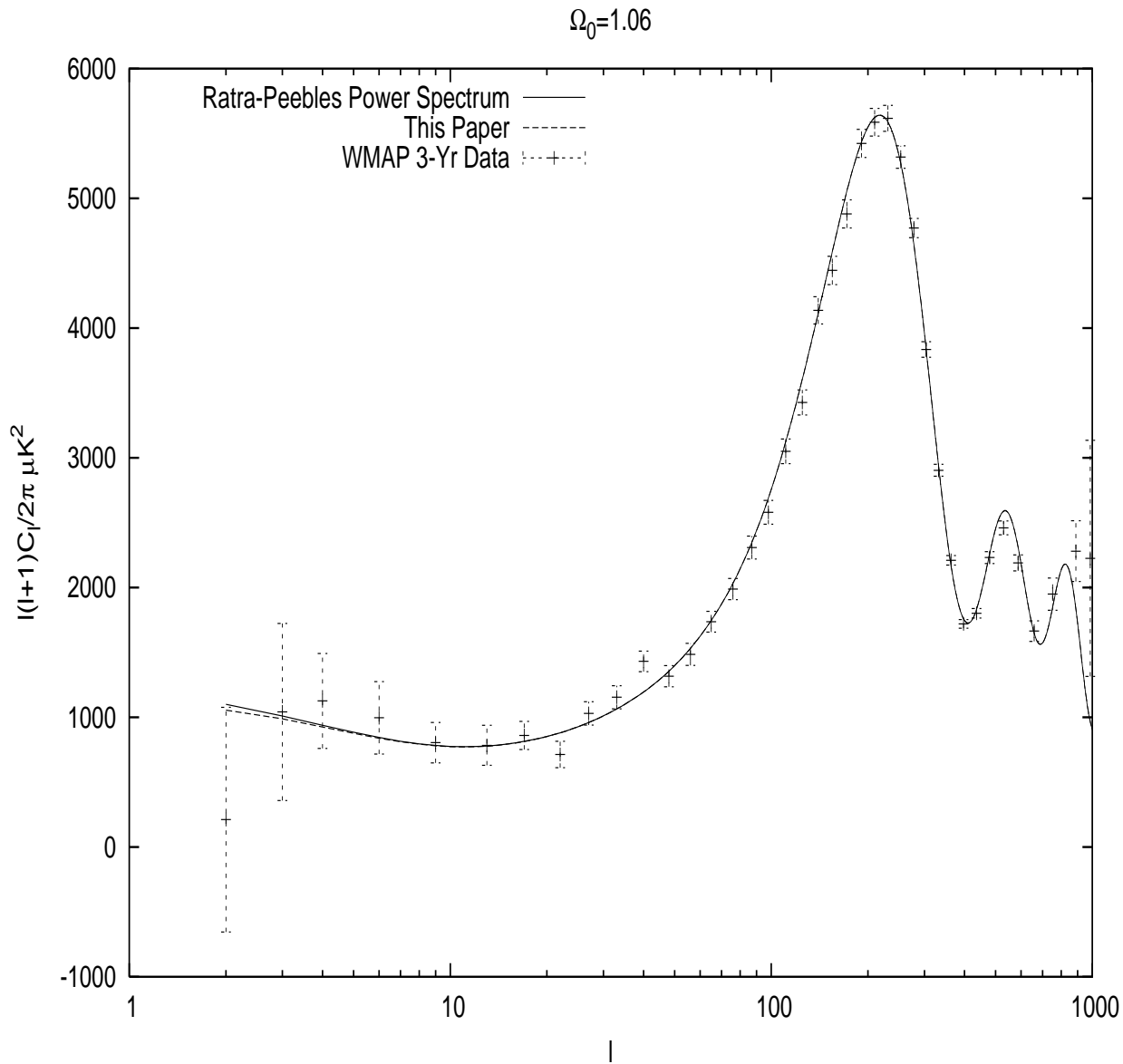}
\end{center}
\caption{Comparison of temperature anisotropy with the Ratra-Peebles power spectrum (\ref{ratra}) and the power spectrum (\ref{closed2}) derived assuming a Bunch-Davies vacuum. The temperature anisotropy has been calculated for a closed universe with $\Omega_0=1.06$ }\label{Fig1}
\end{figure}
\begin{figure}[htb]
\begin{center}
\includegraphics[width=16cm, height=16.5cm,angle=-90]{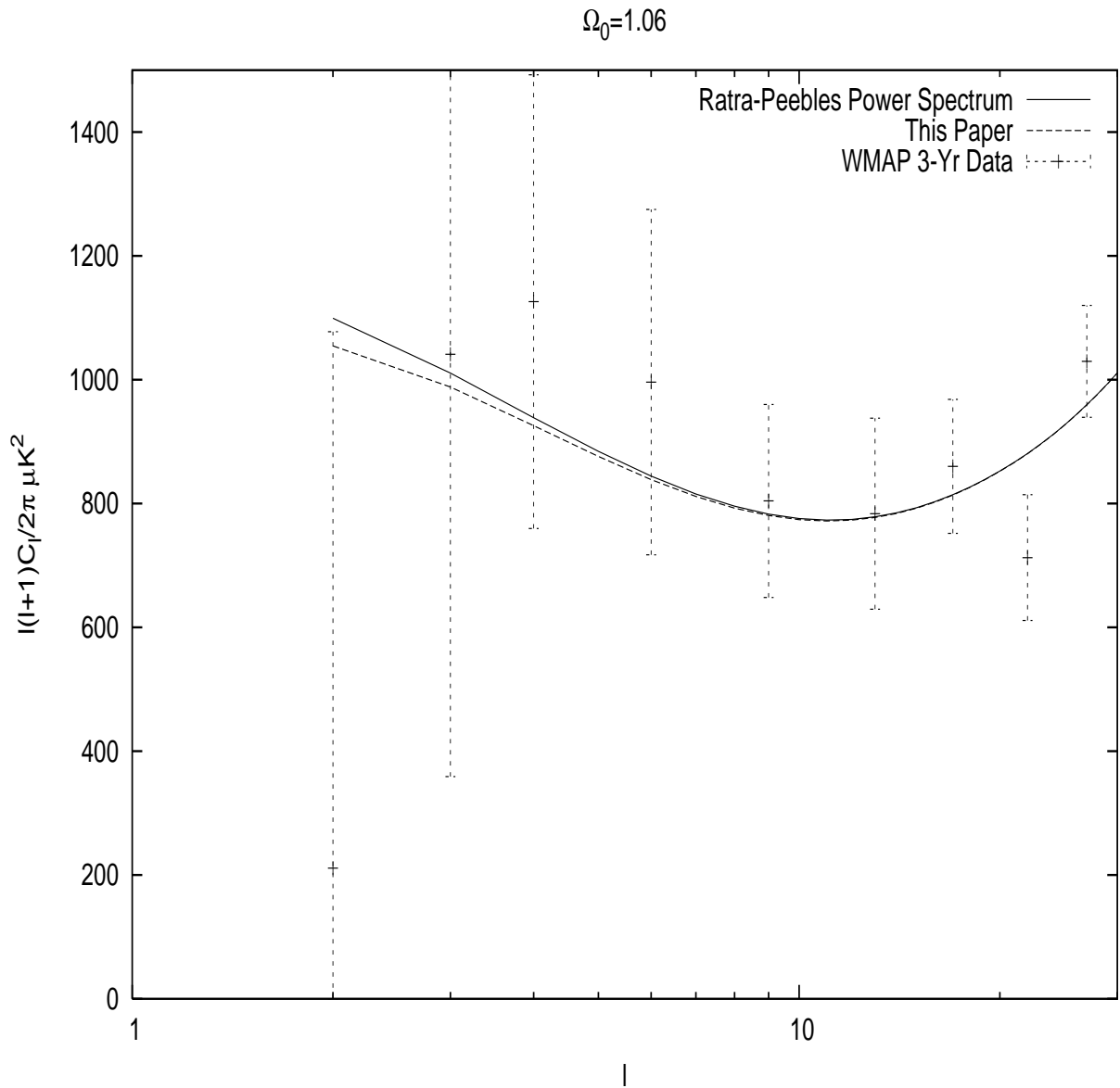}
\end{center}
\caption{Comparison of temperature anisotropy with the Ratra-Peebles power spectrum (\ref{ratra}) and the power spectrum (\ref{closed2}) at low values of $l$ for a closed universe with $\Omega_0=1.06$. }\label{Fig2}
\end{figure}
\begin{figure}[htb]
\begin{center}
\includegraphics[width=16cm, height=16.5cm,angle=-90]{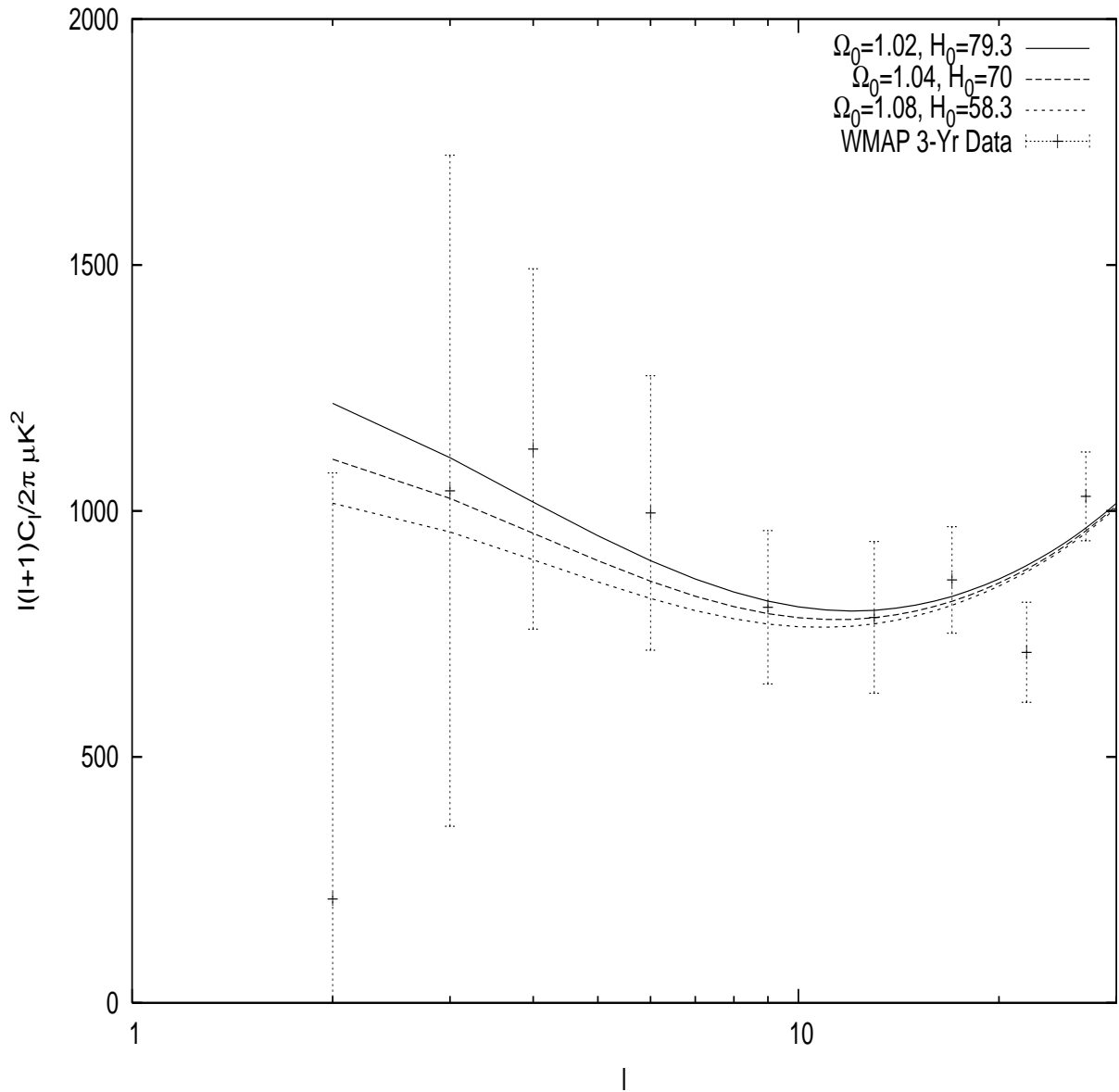}
\end{center}
\caption{Suppression  of quadrupole temperature anisotropy with increasing spatial curvature from the power spectrum (\ref{closed2}). }\label{Fig3}
\end{figure}
\begin{figure}[htb]
\includegraphics[width=11cm, height=11cm,angle=-90]{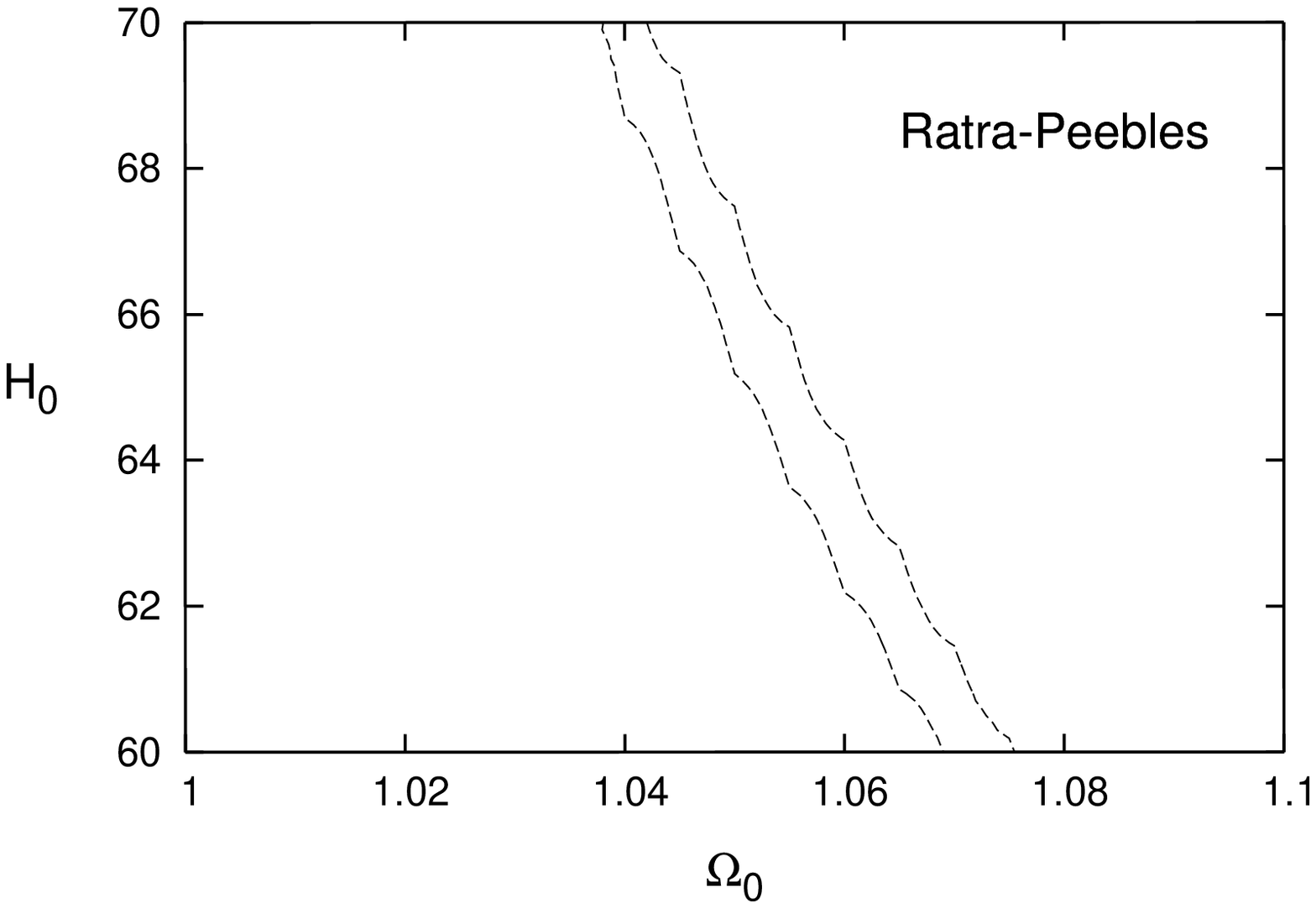}
\includegraphics[width=11cm, height=11cm,angle=-90]{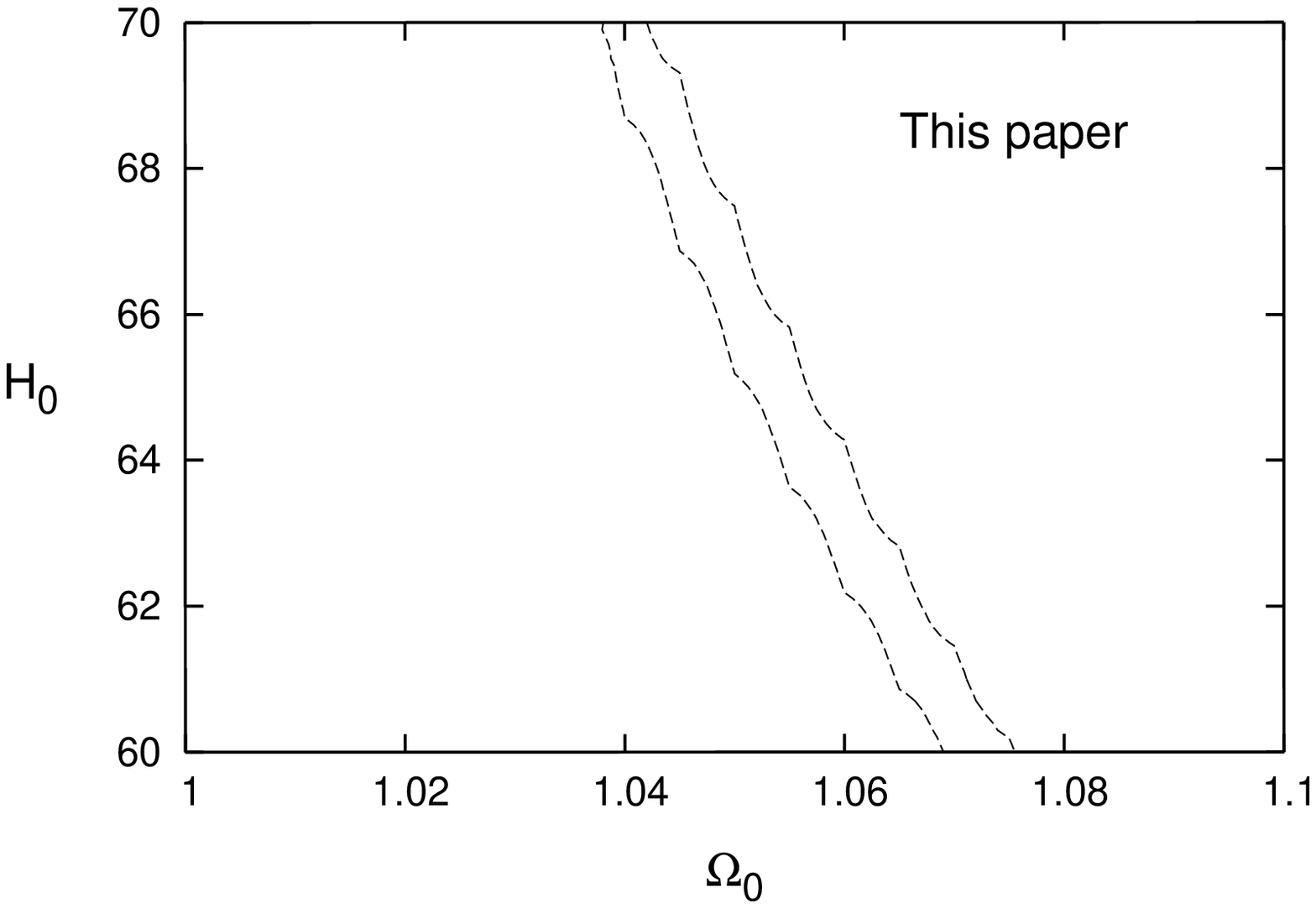}
\caption{The allowed parameter space for $\Omega_0$ and Hubble parameter $H_0$ (in units $Km/sec\,/Mpc$ ) at $90\% C.L$ from WMAP3 data. There is no discernible difference in the parameter space when one assumes the Ratra-Peebles form of power spectrum (\ref{ratra}) or the form (\ref{closed2}) calculated in this paper. }\label{Fig4}
\end{figure}
There are  first principle calculations of power spectrum in a non-flat
inflationary universe \cite{ratra}, \cite{stewart}, \cite{lyth2}, \cite{allen},
\cite{starobinsky}, \cite{sasaki}, \cite{bucher} and \cite{turok}.
Our results for the primordial power spectra for both the closed
and open pre-inflation universe cases differ in some details from
these earlier papers because of differences in the way we have
implemented the initial conditions. Our results of the primordial
power spectrum have been derived assuming that the vacuum state in
the infinite past was the Bunch-Davies vacuum and we have
evaluated the primordial power spectrum at horizon crossing of the
perturbation modes.

The power spectrum obtained by  Ratra and Peebles \cite{ratra} and
Lyth and Stewart \cite{stewart} for the open universe case,
obtained by assuming  conformal boundary condition for the initial
state at $\eta \rightarrow -\infty$ is
\begin{equation}
P_{\cal R}(\beta)=\frac{H_\lambda^4}{2 \pi^2 \dot \phi^2}
\frac{1}{\left(1+\frac{|K|}{\beta^2}\right)}. \label{open3}
\end{equation}
This is sometime written in the form
 \begin{equation}
  \beta^{-3} P_{\cal R}(\beta) \propto  \frac{1}{\beta( \beta^2 +K)}
  \equiv \frac{1}{\beta(\beta^2+1)}.
  \label{ratra}
 \end{equation}
Bucher, Goldhaber and Turok \cite{bucher} consider an open
universe with a tunneling solution and assume that the initial
states annihilate the Bunch-Davies vacuum and obtain a power
spectrum ,
 \begin{equation}
  P_{\cal R}(\beta)=\frac{H_\lambda^4}{2 \pi^2 \dot
  \phi^2} \frac{1}{
  \left(1+\frac{|K|}{\beta^2}\right) }\, \coth\left[
  \frac{\pi \beta}{\sqrt{|K|}}\right].
  \label{open4}
 \end{equation}
In our paper we also assume a Bunch-Davies vacuum but we consider
the standard slow roll inflation model, where the expansion was
dominated by the curvature term prior to inflation, and evaluate
the power spectrum at the horizon exit $ a_*\,H(a_*)=\beta$. In
our solution for the power spectrum of the open universe case
(\ref{open1}) we have a factor of $ 1/(1-|K|/\beta^2)$ instead of
$\coth(\pi \beta/\sqrt{|K|})$ of (\ref{open4}). All three
solutions for the power spectrum (\ref{open1}), (\ref{open3}) and
(\ref{open4}) agree in the limit of small curvature
$|K|/\beta^2\rightarrow 0$.

The experimental bounds on the total density of the universe from
a combination of WMAP, LSS and HST supernovae observations is
$\Omega_0=1.06 \pm 0.02$ \cite{Dunkley}. This implies that the
curvature
\begin{equation}
K=(\Omega-1)\,H_0^2 a_0^2 = (0.06 \pm 0.02)\,H_0^2 a_0^2.
\end{equation}
If one uses the Ratra-Peebles form of the power spectrum for the
closed universe
\begin{equation}
P_{\cal R}(\beta)=\frac{H_\lambda^4}{2 \pi^2 \dot
\phi^2} \frac{1}{
\left(1-\frac{K}{\beta^2}\right) }
\end{equation}
we see that for  perturbations of the horizon size $\beta \simeq
H_0 a_0$, the power spectrum is suppressed by up to $8 \%$
(compared to the flat universe). On the other hand if one uses the
power spectrum (\ref{closed1}) for the closed universe derived in
this paper the suppression of large scale power can be as large as
$16\%$.

In principle the choice of power spectrum used as an  input (in
numerical programmes like CAMB \cite{CAMB} and CMBFAST
\cite{CMBFAST}) will affect the  determination of cosmological
parameters like $\Omega$, $H_0$, $n_s$ etc from the CMB data. In
Fig. \ref{Fig1} we show the temperature anisotropy for a closed
universe with $\Omega_0=1.06$ calculated using the power spectrum
(\ref{closed2}) (dashed line) and the temperature anisotropy
calculated using the Ratra-Peebles power spectrum (\ref{ratra})
(solid line). We modified the CAMB program to determine the
temperature anisotropy spectrum and we have taken the best fit
values of all other parameters like $n_s, h, \tau $ etc. We find
that there is some difference between the two close to $l=2$ but
essentially no difference at large $l$. The difference at lower
$l$ is highlighted in Fig. \ref{Fig2} where we have shown the same
plot as in Fig. \ref{Fig1}, but only for the low $l$ values. We
see that the temperature anisotropy calculated using
(\ref{closed2}) fits the WMAP quadrupole anisotropy data slightly
better than the one calculated using the Ratra-Peebles form.

In Fig. \ref{Fig3} we show that in the case of closed universe for
larger values of $ \Omega_0$ the quadrupole anisotropy is even
more suppressed and fits the WMAP data better using the closed
universe power spectrum (\ref{closed2}).  This supports the idea
proposed in \cite{efsth} that a positive spatial curvature should
suppress the power at low $l$.

In Fig. \ref{Fig4} we show the allowed parameter space of the
Hubble parameter and curvature from the WMAP data. We have used
the power spectrum of this paper (\ref{closed2}) and the
Ratra-Peebles form (\ref{ratra}) to calculate the theoretical
prediction for the temperature anisotropy using CAMB.
Marginalising all other parameters we plot the  allowed values of
$H_0$ and $\Omega_0$ at $90 \% C.L$. Since the theoretical
prediction from the two power spectra match closely except at low
$l$, the chi-square from the two differs only in the second
decimal place and the allowed parameter space from the two power
spectra are almost identical as shown in Fig. \ref{Fig4}.

\section{Conclusion}
At the beginning of inflation the curvature $\Omega-1 $ is
expected to be of order one. By the time perturbations of our
horizon size exit the inflation horizon, the curvature drops to
$\Omega_0-1$ which is the present value. A non-zero observation of
the curvature will tell us whether the universe prior to inflation
was open or closed (even though it is almost flat now) and put
constraints on the number of extra e-foldings that must have
occurred beyond the minimum number needed to solve the horizon
problem. Spatial curvature is a threshold effect which can give us
information on the pre-inflation universe from observations of the
CMB anisotropy at large angles, similar to the effect of a
possible pre-inflation thermal era \cite{temp,gravity}.
>From the power spectrum of
the closed (\ref{closed1}) and open inflation (\ref{open1})
cases we see that if $K >0$, power
is suppressed at large angles and if $K<0$ power is enhanced at
large angles. The WMAP observation of a suppression of the quadrupole temperature anisotropy supports inflation in a closed universe \cite{efsth}.
The spatial curvature is determined in the CMB anisotropy mainly from the
angular size of the acoustic horizon. The determination of the spatial curvature from the WMAP data is not substantially
affected by the choice of the boundary condition used for the determination of the primordial power spectrum in a curved
inflationary universe.

\section{acknowledgements} The work of EM and
GZ is partially supported by Spanish grant  FPA-2005-05904, by
Catalan DURSI grant 2005SGR00916, and by EU 6th Framework Marie
Curie Research and Training network "UniverseNet"
(MRTN-CT-2006-035863). GZ is also supported by DURSI under grant
2003FI00138. SM is supported by the Generalitat de Catalunya under
the program PIV1-2005.

\end{document}